\documentclass{svmult}

\usepackage{graphicx}    

\def\beq{\begin{equation}}
\def\eeq{\end{equation}}

\def\bea{\begin{eqnarray}}
\def\eea{\end{eqnarray}}
\def\ba{\begin{array}}                  
\def\ea{\end{array}}

\newcommand{\pa}{\partial}

\def\L{\Lambda}
\def\m{\mu}

\def\p{\pi}
\def\th{\theta}

\def\s{\sigma}

\def\QEDp{{{\rm{QED}}_+}}
\def\QEDm{{{\rm{QED}}_-}}
\def\SW{{{\mbox {\small {SW}}}}}
\def\P*{{\Psi^*}}
\def\PL*{{\Psi_{\!L}^{*}}}
\def\PR*{{\Psi_{\!R}^{*}}}
\def\p*{{\psi^*}}
\def\lh{{\widehat \Lambda}}

\def\*{\star}
\def\m*{{\star^{\mbox{${_{^{{\!{o{^{_{^{_{\!\!}}}}}p}}}}}$}}}}

\def\Psih{\widehat \Psi}

\def\Fh{\widehat F}
\def\Ah{\widehat A}

\def\fmslashmio{\raisebox{0,5pt}{$\,\slash$} \hspace{- 7 pt} }

\def\Sh{\widehat S}
\def\ell{{{\it l}}}
\def\Tcal{{\cal{T}}}
\def\Tr{{\rm{Tr}}}
\def\Lambdah{\widehat{\Lambda}}
\def\th{\theta}

\def\sma#1{\mbox{\footnotesize #1}}

\def\sk{\vskip .4cm}

\def\stella[#1,#2]{[#1\stackrel{{\displaystyle\star}}{,}#2]}

\begin{document}

\title*{Non-Commutative GUTs, Standard Model and 
$C,P,T$ properties from Seiberg-Witten map}
\titlerunning{NCGUTs, NCSM and C,P,T}
\author{Paolo Aschieri}
\institute{Dipartimento di Scienze e Tecnologie
Avanzate, Universit\'a del Piemonte\\ Orientale, and INFN, Piazza
Ambrosoli 5, I-15100,  Alessandria, Italy\\
Max-Planck-Institut f\"{u}r Physik F\"{o}hringer Ring 6,
D-80805 M\"{u}nchen\\
Sektion Physik,
Universit\"{a}t M\"{u}nchen Theresienstra\ss e 37, D-80333
M\"{u}nchen
\texttt{aschieri@theorie.physik.uni-muenchen.de}}

\maketitle

\begin{abstract}
\noindent
Noncommutative generalizations of Yang-Mills theories using
Seiberg-Witten map are in general not unique. We study these
ambiguities and see that 
$SO(10)$ GUT, at first order in the noncommutativity 
parameter $\th$, is unique and therefore is a truly 
unified theory, while $SU(5)$ is
not. We then present the noncommutative Standard Model
compatible with $SO(10)$ GUT.
We next study the reality, hermiticity and $C,P,T$ properties of 
the Seiberg-Witten map and of these noncommutative actions at all
orders in $\theta$.
This allows to compare the Standard Model discussed in \cite{SM} with the 
present GUT inspired one.
\end{abstract}


\section{Introduction}

There are different examples of noncommutative theories,
we here concentrate on the case where noncommutativity
is described by a constant parameter $\theta^{\mu\nu}$.
The commutation relations among the coordinates read
$\stella[x^\mu,x^\nu]\equiv x^\mu\*x^\nu-x^\nu\* x^\mu=i\th^{\mu\nu}$,
where the star product between functions $f,g$ 
is given by $f\*g=f {_{\,}}e^{{i\over 2}{\th}^{\mu\nu}
{\stackrel{\leftarrow}{\partial}_{\!\mu}}{\stackrel{\rightarrow}
{\partial}_{\!\nu}}} {_{\,}}g_{_{_{_{_{}}}}}$.
We do not claim that spacetime has exactly this noncommutativity, 
rather we are interested in investigating a mathematically
sound gauge theory based on this easiest noncommutative structure.
General aspects of this noncommutative theory will then probably 
be in common with more refined choices of $\th$. In particular the 
choice $\theta^{\mu\nu}=$ constant breakes the Lorentz group in a 
spontaneous 
way; in a bigger theory we would like to consider $\th^{\mu\nu}$ 
(or the related $B$ field)  dynamical and not frozen 
to a constant value, thus recovering Lorentz covariance. 
One can also consider gauge theories with $\th$ nondynamical
but frozen to a particular nonconstant value, 
linear in the coordinates, such that one has a (kappa) deformed 
Poincar\'e symmetry, see \cite{Wess}.

Using Seiberg-Witten map \cite{SW}, 
that relates commutative 
gauge fields to noncommutative ones in such a way that commutative 
gauge transformations are mapped in NC gauge transformations, one can
construct NC gauge theories with arbirary gauge groups 
\cite{Madore, Jurco1}.
These theories are invariant under both commutative and noncommutative 
gauge transformations. Along these lines noncommutative 
generalizations of the 
standard model and GUT theories have been studied 
\cite{SM,A}.  
The SW map and the $\*$ product 
allow us to expand these noncommutative actions order by order in $\th$ 
and to express them in terms of ordinary commutative fields
so that one can then study the physics properties of these
$\th$-expanded commutative actions, see for ex. \cite{astron}.

It turns out that given a commutative YM theory, 
SW map and commutative/noncommutative gauge invariance 
are in general not enough in order to single out a unique 
noncommutative generalization of the
original YM theory. One can follow different criteria in order to
select a specific noncommutative generalization. We here focus on a
classical analysis, in particular imposing the constraint that 
the noncommutative generalization of the Standard Model should 
be compatible with noncommutative GUT theories. 
Another issue would be to single out a noncommutative SM or GUT that 
is well behaved at the quantum level. 
We refer to the problems relative to 
renormalization, see for ex. \cite{Wulkenhaar2}. On the other hand 
chiral gauge anomalies are absent in these models
\cite{Brandt}.

\sk
In this talk, following \cite{A}, we present a general study of the ambiguities
that 
appear when constructing NCYM theories. We then see that at first order in 
$\th$ there is no ambiguity in $SO(10)$ NCYM theory. In particular 
no triple gauge bosons coupling  of the kind  $\th FFF$
 is present.
We further study the noncommutative SM compatible with $SO(10)$.

We next study the reality, hermiticity, 
charge conjugation, parity and time reversal properties of 
the SW map and of $\th$-expanded NCYM theories. 
This constraints the possible freedom in the choice of a
``good'' SW map.
In \cite{Sheikh-Jabbari:2000vi} the $C,P,T$ properties of NCQED
were studied assuming the usual $C,P$ and $T$ transformations
also for noncommutative fields. 
We here  show that the usual
$C,P,T$ transformation on commutative spinors and nonabelian
gauge potentials imply, via SW map, the same $C,P,T$
transformations for the noncommutative spinors and gauge potentials.
We also see that $CPT$ is always a  symmetry of noncommutative
actions. In \cite{Chai} $CPT$ is studied more axiomatically.

The reality property of the SW map is then used to 
analyze the difference between the SM in \cite{SM} and the GUT inspired 
SM proposed here. 
It is a basic one, and can be studied also in a QED model. 
While in \cite{SM}, and in general in the literature,
left and right handed components of a noncommutative spinor field are 
built with the same SW map, we here use and advocate a different
choice: if noncommutative left handed fermions are built with the 
$+\th$ SW map then their right handed companions should be built with 
the $-\th$ SW map; this implies that both noncommutative $\psi_L$ and
$\psi^{\scriptscriptstyle C}_{~L}\equiv
-i\sigma_2{_{_{\,}}}{\psi_{_{\!}R}}^{^{_{_{_{\scriptstyle *}}}}}$
are built with the $+\th$ SW map.
In other words, with this choice, noncommutativity does not
distinguish 
between a left handed fermion and a left handed antifermion, but 
does distinguish between fermions with different chirality.
This appears to be the only choice compatible with GUT theories.

\section{Seiberg-Witten map and NC particle models}
Consider an ordinary ``commutative'' YM action with  gauge group $G$, 
and one fermion multiplet,
$\,\int d^4x \,{-1\,\over 2 g^2}^{\,} 
\Tr(F_{\mu\nu}F^{\mu\nu})  + 
\overline{\Psi}  i
{\fmslashmio D} \Psi \,.$
This action is gauge invariant under 
$
\delta\Psi=i\rho_\Psi(\Lambda)\Psi\label{psi}
$
where $\rho_{\Psi}$ is the representation of $G$  determined by $\Psi$.
Following \cite{Jurco1}
the noncommutative generalization of 
this action is given by 
\beq
\Sh=\int d^4x \,{-1\,\over 2 g^2}^{\,} 
T^{\!}r(\Fh_{\mu\nu}\*\Fh^{\mu\nu})  + 
\overline{\widehat \Psi} \star i
\widehat{\fmslashmio D} \widehat \Psi \label{Action1multiplet}
\eeq
where the noncommutative field strength $\Fh$ is defined by
$
\widehat F_{\mu\nu}  = \pa_\mu\widehat A_\nu 
- \pa_\nu\widehat A_\mu  -i\s[\widehat A_\mu,\widehat A_\nu]\,.
$
The covariant derivative is given by
\beq
\widehat D_\mu \widehat\Psi = \pa_\mu \widehat\Psi 
- i \rho_\Psi(\widehat A_\mu)\star\widehat\Psi\,.
\label{covder}
\eeq
The action (\ref{Action1multiplet}) is invariant under the 
noncommutative gauge transformations 
\beq
\hat\delta\widehat\Psi = i \rho_\Psi(\widehat\Lambda) \star
\widehat\Psi ~~~,~
~~\hat\delta \widehat A_\mu = \pa_\mu\widehat\Lambda 
+ i\s[\widehat\Lambda,\widehat A_\mu]~.
\eeq

The fields $\Ah$, $\Psih$ and $\lh$ are functions of the commutative 
fields $A,\Psi, \Lambda$ and the noncommutativity parameter $\th$
via the SW map \cite{SW}. At first order in $\th$ we have
\begin{eqnarray}
\widehat A_\xi[A,\th] & = & A_\xi 
+ \frac{1}{4} \theta^{\mu\nu}\{A_\nu,\pa_\mu A_\xi\} + 
\frac{1}{4} \theta^{\mu\nu}\{F_{\mu\xi},A_\nu\} +  \mathcal{O}(\theta^2)
\label{SWA}\\
\widehat\Lambda[\Lambda, A, \th]  &=&  \Lambda 
+ \frac{1}{4} \theta^{\mu\nu}\{\pa_\mu \Lambda , A_\nu\}+ \mathcal{O}(\theta^2)\label{SWLambda}~\\
\widehat \Psi[\Psi,A,\th] & = & \Psi 
+ \frac{1}{2} \theta^{\mu\nu}\rho_\Psi(A_\nu)\pa_\mu\Psi
+\frac{i}{8}\theta^{\mu\nu}[\rho_\Psi(A_\mu), \rho_\Psi(A_\nu)] \Psi + \mathcal{O}(\theta^2)
 \label{SWPsi}
\end{eqnarray}
In terms of the commutative fields 
the action (\ref{Action1multiplet}) is also invariant
under the ordinary gauge transformation 
$\delta A_\mu = \pa_\mu\Lambda  + i[\Lambda, A_\mu]$,
$\delta\Psi=i\rho_\Psi(\Lambda)\Psi$.
\sk
In (\ref{Action1multiplet}) the information on the gauge group $G$ 
is through the dependence of the noncommutative fields on the 
commutative ones. The commutative
gauge potential $A$ and gauge parameter $\Lambda$ are valued in the
$G$  Lie algebra, $A=A^aT^a, \L =\L^aT^a$; and from 
(\ref{SWA}), (\ref{SWLambda})
it follows that $\Ah$ and $\Lambdah$ 
are valued in the universal enveloping 
algebra of the $G$ Lie algebra. 
However, due to the SW map, the degrees of freedom of $\Ah$ are the same as that of $A$.
Similarly to $\Ah$, also $\Fh$ is valued in the universal enveloping 
algebra of $G$.
Now expression (\ref{Action1multiplet}) is ambiguous because 
in  $T^{\!}r(\Fh_{\mu\nu}\*\Fh^{\mu\nu})$ we have not specified 
the representation $\rho(T^a)$. 
We can render explicit the ambiguity in (\ref{Action1multiplet}) 
by writing
\beq
{1\over g^2}T^{\!}r(\Fh_{\mu\nu}\*\Fh^{\mu\nu})=
\sum_\rho c_\rho\Tr(\rho(\Fh_{\mu\nu})\*\rho(\Fh^{\mu\nu}))
\label{ambiguity}
\eeq
where the sum is extended over all unitary irreducible and
inequivalent representations $\rho$ of $G$. 
The real coefficients $c_\rho$ parametrize the ambiguity in 
(\ref{ambiguity}). They are constrained by 
requiring that in the commutative limit, 
$\th\rightarrow 0$, (\ref{ambiguity}) becomes the correctly normalized
commutative gauge kinetic term.

The ambiguity (\ref{ambiguity}) in the action (\ref{Action1multiplet})
can also be studied by expanding (\ref{ambiguity}) in terms of the
commutative fields $\Psi,A,F$.  At first order in $\th$ we have
\bea
\widehat{S}_{gauge}&=&
-\frac{1}{4g^2}\int \! d^4x 
\sum_{a=1}^{{\mbox{{\tiny{dim }}}}{\!G}}
F^a_{\mu \nu}F^{a\,\mu \nu}_{\!}\nonumber\\
& &\,+\,(_{\,}{\mbox{$\sum_\rho$} c_\rho D_\rho^{abc}}) \,\,\,
{\theta^{\mu \nu}\over 4}\!\int \!d^4x^{\,\,\,} {1\over 4}F^a_{\mu \nu}
F^b_{\rho \sigma} F^{c\,\rho \sigma}\, -\,F^a_{\mu \rho} F^b_{\nu \sigma}
F^{c\,\rho\sigma}~~~~~~~~~
\label{traccia}
\eea
where 
\beq
{1\over 2}D_\rho^{abc}\equiv\Tr(\rho(T^a)\{\rho(T^b),\rho(T^c)\})
= {\cal A(\rho)}\Tr(t^a\{t^b,t^c\})\equiv
{1\over 2}{\cal A(\rho)}d^{abc}\label{Dterm}~.
\eeq
Here $t^a$ denotes the fundamental representation, and we are
using that the completely symmetric $D_\rho^{abc}$ tensor in the
representation $\rho$ is proportional to the $d^{abc}$ one defined
by the fundamental representation. In particular 
for all simple Lie groups, except $SU(N)$ with $N\geq 3$, 
we have $D_\rho^{abc}=0$
for any representation $\rho$.
Thus from  (\ref{traccia}) we see that at first order in $\th$  
the ambiguity (\ref{ambiguity}) is present just for  $SU(N)$ Lie
groups.
\sk
Among the possible representations that one can choose in 
(\ref{ambiguity}) there are two natural ones. 
The fermion representation and the adjoint representation. 
The adjoint representation is particularly appealing if we just have a pure
gauge action, then, since only the structure constants appear in the
commutative gauge kinetic term $\sum_a F_{\mu\nu}^aF^{a\,\mu\nu}$, 
a possible choice is indeed to consider only the
adjoint representation. This is a minimal choice in the sense that in
this case only structure constants enter (\ref{ambiguity}).
(It can be shown \cite{A} that in this case the gauge action is even
in $\th$).
If we also have matter fields then from (\ref{covder})
we see that we must consider the particle representation  
$\rho_\Psi$ given by the multiplet $\Psi$ (and inherited by $\Psih$). 
In (\ref{ambiguity}) one could then make the minimal choice
of selecting just the $\rho_\Psi$ representation.

Along the lines of the above NCYM theories framework we now examine
the $SO(10)$, the $SU(5)$  and the Standard Model noncommutative 
gauge theories.
\sk

\noindent{\bf Noncommutative \mbox{\boldmath ${ SO(10)}~$}}
We consider only one fermion generation: the $16$-dimensional spinor representation of
$SO(10)$ usually denoted $16^+$ (no relevant new effects appear
considering all three families).
We write the left handed multiplet as 
\beq
\Psi_L^+=(u^i,d^i\,,\,-u^{{\scriptscriptstyle C}}_i\,,\,
d^{{\scriptscriptstyle   C}}_i\,,\,\nu,e^-\,,\,e^+\,,\,
-\nu^{\scriptscriptstyle  C})_L
\label{ml}
\eeq
where $i$ is the $SU(3)$ color index and 
$\nu^{\scriptscriptstyle  C}_{~L}=
-i\sigma_2{_{_{\,}}}{\nu_{_{\!}R}}^{^{_{_{_{\scriptstyle *}}}}}$
is the charge conjugate of the neutrino particle $\nu_R$ 
(not present in the Standard Model).
The gauge and fermion sector of noncommutative $SO(10)$ is then simply
obtained by replacing  $\Psih$ with $\Psih^+_L$ in (\ref{Action1multiplet}).
Notice that no linear term in $\th$, i.e. no cubic term in $F$ can appear. 
This is so because $SO(10)$ is anomaly free: $D_\rho^{abc}=0$ forall $\rho$.
In other words, at first order in $\th$, noncommutative $SO(10)$ gauge
theory is unique. 
\sk
\noindent{\bf Noncommutative \mbox{\boldmath ${ SU(5)}~$}}
 The fermionic sector of $SU(5)$ has the
${\psi^{\scriptscriptstyle C}}_L$ multiplet 
that transforms in the $\overline{5}$ of $SU(5)$ and the 
$\chi_L$ multiplet that
transforms according to the $10$ of $SU(5)$. 
In this case we expect that the adjoint, the ${\overline 5}$
and the $10$ representations enter in (\ref{ambiguity}). 
In principle one can consider the coefficients
$c_{\overline 5}\not=c_{10}$, i.e. while the 
$({\psi^{\scriptscriptstyle C}}_L,\chi_L)$ 
fermion rep.
is $\overline 5\oplus 10$, in (\ref{ambiguity}) the weights
$c_\rho$ of the  ${\overline 5}$ and the $10$ can possibly 
be not the same.
It turns out that only if $c_{\overline 5}\not=c_{10}$ then  
$\sum_\rho c_\rho D_\rho^{abc}\not=0$ in (\ref{traccia}). 
We see that, already at first order in the noncommutativity 
parameter $\th$,  noncommutative $SU(5)$ gauge theory 
is not uniquely determined by the gauge coupling constant $g$, 
but also by the value of $\sum_\rho c_\rho D_\rho^{abc}$.
Thus $SU(5)$ is \emph{not} a truly unified
theory in a noncommutative setting.
It is tempting to set $c_{\overline 5}=c_{10}$ so that 
exactly the fermion representation ${\overline 5}\oplus 10$ enters 
(\ref{traccia}). We then have $\sum_\rho c_\rho D_\rho^{abc}=0$,
(however this relation is not protected by symmetries).

\sk 
\noindent {\bf (GUT inspired) Noncommutative Standard Model $~$}
One proceeds similarly for the SM gauge group. The full ambiguity of
the gauge kinetic term is given in \cite{A}.
About the fermion kinetic term, the fermion vector  
${\widehat \Psi_L}$ is constructed from 
$\Psi_L=(u^i,d^i\,,\,-u^{\scriptscriptstyle C}_i\,,
\,d^{\scriptscriptstyle C}_i\,,\,\nu,e^-\,,\,e^+)_L$.
The  covariant derivative is as in (\ref{covder}), 
with $\Psi\rightarrow \Psi_L$ and with $A_\mu=A_\mu^A\Tcal^A$,
where $\{\Tcal^A\}=\{ Y,T^a_L,T^\ell_S\}$ are the generators of 
$U(1)\otimes SU(2)\otimes SU(3)$.
The fermion kinetic term is then as in  (\ref{Action1multiplet}).
This Standard Model is built using only left handed fermions and
antifermions. We call it GUT inspired because its noncommutative
structure can be embedded in $SO(10)$ GUT. Indeed $\Psi_L$ and 
$\Psi_L^+$  differ just by the extra neutrino 
$\nu^{\scriptscriptstyle  C}_{~L}=
-i\sigma_2{_{_{\,}}}{\nu_{_{\!}R}}^{^{_{_{_{\scriptstyle *}}}}}$;
moreover under an infinitesimal gauge transformation $\lh$,
all fermions in $\Psi_L$ transform with $\lh$ on the left.
This GUT inspired Standard Model differs from the one considered 
in \cite{SM}; indeed here we started from the chiral vector
$\Psi_L$, while there the vector $\Psi'=(u^i_L,d^i_L\,,\,u^i_R\,d^i_R\,,\,\nu_L,e^-_L\,,\,e^-_R)$
is considered. In the commutative case $\int \overline{\Psi'} 
{\fmslashmio D} \Psi'=\int \overline{\Psi_L}  
{\fmslashmio D} \Psi_L$ but in the noncommutative case (see later)
this is no more true:  $\int \overline{\widehat{\Psi'\,}} \*
{\widehat{\fmslashmio D}} \widehat{\Psi'\,}\not=
\int \overline{\widehat{\Psi_{\!L}}} \*
{\widehat{\fmslashmio D}}_{_{\,}} \widehat{\Psi_{\!L}}\,$;
if we change  $\th$ into $-\th$  in the right handed sector of 
$\int \overline{\widehat{\Psi'\,}} \*
{\widehat{\fmslashmio D}} \widehat{\Psi'\,}$, 
then the two expressions coincide.

Finally it is a natural choice to consider in the SM gauge kinetic term
only the adjoint rep. and the fermion rep.,
we then have  that at first order in $\theta$ there are no
modifications to the SM gauge kinetic term.
This is so because the fermion rep. is anomaly free:
$D_{\rho_{{_{\rm{fermion}}}}}^{A^{\,_{\!\!}}A'\!A''}=0$ 
and because for $U(1)$ the adjoint rep. is trivial.

\sk
\noindent {\bf Higgs Sectors$~$}
While the noncommutative Higgs kinetic and potential
terms are given by 
\beq
(\widehat D_\mu \widehat \phi)^\dagger
\star \widehat D^\mu \widehat \phi           
\,+\, \mu^2 {\widehat{\phi}}^{\,\dagger} \star  \widehat \phi - \lambda\,
\widehat{\phi}^{\,\dagger} \star  \widehat{\phi}
\star
{\widehat \phi}^{\,\dagger} \star {\widehat \phi}\,   ~,
\eeq
a noncommutative version of the SM and GUT Yukawa
terms is not straighforward and requires the introduction of the
hybrid Seiberg-Witten 
maps  ${\widehat{{\quad}}}^{^{_{_{_{\!\!H}}}}}\!$ 
on fermions.
A typical noncommutative Yukawa term then reads
\beq
\widehat{\phi}^{{{\,}}\dagger} \*
\widehat{L_L}^{^{_{_{_{\!\!H}}}}}\*
\widehat{e_R^*}+~herm. ~con_{\!}j.
\label{hsw}
\eeq
where 
$L_L^{}=\left({^{^{^{\!}}}}{_{_{_{\!\!}}}}\right.
{}^{^{\mbox{\sma{$\nu^{}_L $}}}}_{_{\mbox{\sma{$e^{}_L $}}}}
\left.{^{^{^{\!}}}}{_{_{_{\!\!\!}}}}\right)\,$.  
Under an infinitesimal $U(1)\otimes SU(2)\otimes SU(3)$ 
gauge transformation
$\Lambda$, 
${\widehat{{L_L^{}}}}^{^{_{_{_{\!\!H}}}}}\!$ transforms as
$
\delta\,{\widehat{{L_L^{}}}}^{^{_{_{_{\!\!H}}}}}\!=
i\rho_\phi(\lh)\*{\widehat{{L_L^{}}}}^{^{_{_{_{\!\!H}}}}}\!
-i{\widehat{{L_L^{}}}}^{^{_{_{_{\!\!H}}}}}\!\*
\rho_{e_R^*}(\lh)
$.
We see that in the hybrid SW map $\lh$ appears both on the
left and on the right of the fermions, moreover
the representation of $\lh$
is inherited from the Higgs and fermions that sandwich 
${\widehat{{L_L^{}}}}^{^{_{_{_{\!\!H}}}}}\!$.
The Yukawa term (\ref{hsw}) is thus  invariant under
noncommutative gauge transformations. Of course in the $\th\rightarrow
0$ limit we recover the usual gauge transformation for the leptons.
An explicit formula for the hybrid SW map at first order in $\th$ is in 
\cite{SM, A}.
The Yukawa terms (\ref{hsw}) differ from those studied in \cite{SM}.
There the hybrid SW map is considered on $\phi$, in particular there
${\widehat{\phi}}^{^{^{{{H}}}}}\!$ is not invariant under $SU(3)$
gauge transformations (and this
implies that in \cite{SM} gluons couple directly to the Higgs field).

The Higgs sector in the $SO(10)$ and $SU(5)$ models can be constructed
with similar techniques \cite{A}.

\sk
\section{Hermiticity and reality properties of SW map}
From (\ref{SWA}) we see that if $A$ is hermitian
then $ \Ah$ is also  hermitian.
Actually, to all orders in $\th$, $\Ah$ and $\lh$ 
can be chosen hermitian if $A$ and $\Lambda$ are hermitian. 
Otherwise stated,
SW map can be chosen to be compatible with hermitian conjugation.
Compatibility of SW map with complex conjugation reads,
\beq
\widehat{\P*}={\widehat{\Psi}}^{{^{\,\mbox{\scriptsize{$*$}}}}}
\label{conj}
\eeq
where 
$
\Psih= \SW [\Psi,\rho_\Psi(A), \th]
$
denotes the SW map of $\Psi$ constructed with the representation 
$\rho_\Psi$ of the potential $A$, and the SW map of the complex
conjugate spinor $\P*$ is defined by 
\beq
\widehat{\P*}\equiv \SW [\P*,\rho_\P*(A), - \th]
\label{30}
\eeq
where $\rho_\P*$ is the representation conjugate to 
$\rho_\Psi$\footnote{ 
Given the group element $g=e^{i\Lambda}=e^{i\Lambda^aT^a}$ we have 
$\rho_{\Psi^*}(g)\equiv\overline{\rho_\Psi(g)}$ and, 
since 
$\L^a$,  $A^a$ are real, we have
$\rho_{\P*}(\L)=-\overline{\rho_\Psi(\L)}
{}~,~~\rho_\P*(A)=-\overline{\rho_\Psi(A)}\;
$.}. Notice that in (\ref{30}) the noncommutativity parameter $\theta$
appears with opposite sign w.r.t. the $\theta$ in the SW map of
$\Psi$.
Similarly to (\ref{conj}) we have 
$\widehat{\rho_\P* (A)}=-\overline{{\widehat{\rho_\Psi(A)\,}}}$ and
$\widehat{\rho_\P* (\L)}=-\overline{{\widehat{\rho_\Psi(\L)\,}}}$
where 
$\widehat{\rho_\P* (A)}\equiv\Ah[\rho_\P*(A),-\th]$ and 
$\widehat{\rho_\P* (\L)}\equiv\lh[\rho_\P*(A),\rho_\P*(\L),-\th]$.
{}The proof of (\ref{conj}) relies on showing that the SW differential
equations \cite{SW} (obtained by requiring
that gauge equivalence classes of  the  gauge
theory with noncommutativity 
$\th+\delta\th$, correspond to 
gauge equivalence classes of the gauge
theory with noncommutativity $\th$)
are themselves compatible with complex
conjugation \cite{A}. One proceeds similarly for 
the case of hermitian conjugation.
\sk
\noindent{\bf Noncommutativity and chirality$~$} We can now discuss a further ambiguity of noncommutative gauge theories, and
resolve it by requiring compatibility with grand unified theories.
{}For simplicity we consider noncommutative QED.
Let $\psi$ be a $4$-component Dirac spinor, and
decompose it into its Weil spinors $\psi_L$ and $\psi_R$.
Their charge conjugate spinors are
$\psi^{\;C}_{\!L}=\psi^C_{~L}=-i\sigma_2\psi_R^{\,*}$
and  $\psi^{\;C}_{\!R}=\psi^C_{~R}=i\sigma_2\psi_L^{\,*}$.
Consider 
the  noncommutative left-handed spinor
$\widehat{\psi_L}=\SW [\psi_L,\rho_{\psi_L}(A),\th]~,$
we then have the $\mbox{\boldmath ${\pm}$}\th$ choice 
\beq
\widehat{\psi_R}=\SW [\psi_R,\rho_{\psi_R}(A),\mbox{\boldmath ${\pm}$}
\th]~
\label{choice}
\eeq
for the right handed one.
In the literature the choice $+\th$ is usually
considered so that for the $4$-component Dirac spinor $\psi$ we can
write  $\widehat{\psi}=\SW [\psi,A,\th]$, 
$\delta\widehat{\psi}=i\widehat{\Lambda}\*\widehat{\psi}$.
We here advocate the opposite choice ($-\th$) in (\ref{choice}).
Indeed from  (\ref{conj}) we have that 
$\widehat{\psi_{\!L}^{\;C}}=-i\sigma_2\,{\widehat{\psi_R}}^{\,*}$ and therefore
\beq
\widehat{\psi_R}=\SW [\psi_R,\rho_{\psi_R^{}}(A),-\th]~~~
\Longleftrightarrow~~~
\widehat{\psi_{\!L}^{\;C}}=\SW [{\psi_{\!L}^{\;C}},
\rho_{\psi_{\!L}^{\;C}}(A),+\th]
\label{equiv}
\eeq
so that with the $-\th$ choice in (\ref{choice}), 
both left handed fermions $\widehat{\psi_L}$,
$\widehat{{\psi^{\;C}_{\!L}}}$
are associated with $\th$ while the right handed ones $\widehat{\psi_R}$, 
$\widehat{{\psi^{\;C}_{\!R}}}$ are associated with $-\th$.
In GUT theories we have multiplets of definite chirality
and therefore this is the natural choice to
consider in this setting.

These observations allow us to compare  
QED$_+$ with  $\QEDm$, the two different QED theories obtained with the two
different $\pm\th$ choices (\ref{choice}). This difference 
immediately extends to the fermion kinetic terms 
of nonabelian gauge theories and allows us to compare the 
NCSM discussed in
\cite{SM} with the present GUT compatible one. We have (up to gauge kinetic terms)
$$
S_\QEDp \!
=\int {\widehat{\psi_{\!L}}}^\dagger\* i
\widehat{\fmslashmio D}_{\,} \widehat \psi_{\!L}
\,+\,{\widehat{\psi_{\!R}}}^\dagger\* i
\widehat{\fmslashmio D}_{\,} \widehat \psi_{\!R}
~~~,~~~
S_\QEDm\! =\int {\widehat{\psi_{\!L}}}^\dagger\* i
\widehat{\fmslashmio D}_{\,} \widehat \psi_{\!L}
\,+\,{\widehat{\psi^{\;C}_{\!L}}}^\dagger\* i
\widehat{\fmslashmio D}_{\,} \widehat{\psi^{\;C}_{\!L}}\:~~
$$
where the GUT inspired $\QEDm$ is obtained 
using the left handed spinor
 \(\left({^{^{^{\!}}}}{_{_{_{\!\!}}}}\right.
{}^{^{\mbox{\sma{$_{\psi_{\!L}^{} }$}}}}_{{\mbox{\sma{$^{\psi^C_{\!L} }$}}}}
\left.{^{^{^{\!}}}}{_{_{_{\!\!\!\!}}}}\right)\) 
so that $\widehat{\psi^C_{\!L}}
=\SW[\psi^C_{\!L},\rho_{\psi^C_{\!L}}(A),\th]$. 
Now from 
$\widehat{\psi_{\!L}^{\;C}}=-i\sigma_2\,{\widehat{\psi_R}}^{\,*}$
and from $\sigma$ matrix algebra we have
$\,\int {\widehat{\psi^{\;C}_{\!L}}}^\dagger\* i
\widehat{\fmslashmio D}_{\,} \widehat{\psi^{\;C}_{\!L}}=
\int {\widehat{\psi_{\!R}}}^{\!{{
\mbox{${_{^{{{o{^{_{^{_{\!\!}}}}}p}}}}}$}}}}
{}{^{\mbox{${}^\dagger$}}}\m* i
\widehat{\fmslashmio D}^{\!{{
\mbox{${_{^{{{^{\:{_{\!\!\!}}}}{o{^{_{^{_{\!\!}}}}}p}}}}}$}}}}
_{\,} \widehat{\psi_{\!R}}^{\!{{
\mbox{${_{^{{{o{^{_{^{_{\!\!}}}}}p}}}}}$}}}}
$,
where we have emphasized  that we are using the $-\th$ convention
in the SW map by writing 
${~}\widehat{^{}}^{\!{{\mbox{${_{^{{\,\;{o{^{_{^{_{\!\!}}}}}p}}}}}$}}}}$ 
instead of ${~}\widehat{^{}}{~\,}$. 
We conclude that in order to obtain QED$_-$ from QED$_+$ we just need to 
change $\th$ into $-\th$ in the right handed fermion sector of
QED$_+$.

\section{$C,P,T$ properties of SW map and  of NCYM actions}
Using  compatibility of SW map with complex conjugation and the
 tensorial properties of SW map (i.e. that
 SW map preserves the space-time index)
one can study the properties of SW map with respect to the $C$, $P$
 and $T$ operations. In particular these same expressions as in the
 commutative case holds:
\beq
\widehat{\Psi_{\!L}}^{\:T}={-i\sigma_1\sigma_3
\widehat{\Psi_{\!L}}}~~~,~~~
{\widehat{\Psi_{\!R}}^{\:T}}
={-i\sigma_1\sigma_3\widehat{\Psi_{\!R}}}
~~~,~~~
\Ah_{\,\mu}^{\textstyle{^{\;T}}}=(\Ah_0,-\Ah_i) 
\label{LT}
\eeq
{\vskip -2.5em}
\beq
{\widehat{\Psi_{\!L}}^{\:C\!P}}=i\sigma_2{\widehat{\Psi_{\!L}}}^{*}
      ~~~,~~~
{\widehat{\Psi_{\!R}}^{\:C\!P}}=-i\sigma_2{\widehat{\Psi_{\!R}}}^{*} 
~~~,~~~\Ah_{\,\mu}^{\textstyle{^{\;C\!P}}}
=(-\overline{\Ah_0},\overline{\Ah_i}) 
\label{LCP}
\eeq
where the action of the $P$ and $C$ operators on spinors is given by 
$$
\widehat{\Psi_{\!L}}^{\,P}=
 \SW[\Psi_{\!L}^{\,P},\rho_{\Psi_{_{\!}L}}^{}(A^P),\th^P,\partial^P,i]~~~,~~~
\widehat{\Psi_{\!L}}^{\,C}=
\SW[\Psi_{\!L}^{\,C},{(\rho_{\Psi_{_{\!}L}^{}}(A))^C},
\th^C,\partial,i]~,$$
while the time inversion is given by 
$
{\widehat{\Psi_{\!L}}^{\:T}}
=\SW[\Psi_{\!L}^{{^{^{\,}}}T},
\rho_{\Psi_{\!L}^{{^{^{\,}}}T}}(A^T),
\th^T,\partial^T,-i]\,
$.
In these expressions we have written explicitly the dependence on the 
partial derivatives, and
the imaginary unit $i$ in the last slot marks that the coefficients in
the SW map are in general complex coefficients. 
The $-i$ in the last expression means that we are considering the
complex conjugates of the coefficients in the SW map, this is so
because $T$ is antilinear and multiplicative. 
Relations (\ref{LT}) and (\ref{LCP}) hold provided that $\theta^{\mu\nu}$
transforms under $C,P,T$  as a $U(1)$ field strenght $F_{\mu\nu}$.
If we choose $+\theta$ in (\ref{choice}) then parity and charge
conjugation sepatately assume the same expression as 
in the commutative case.
\sk
Now we discuss the transformations properties
of NCYM actions under $C,P$ and $T$. With the $+\th$
choice (\ref{choice}) we have that NCYM actions are 
invariant under $C,P$ and $T$ iff in the commutative limit they are 
invariant. On the other hand, with the $-\th$ choice
NCYM actions are invariant under $C_{\!}P$ and $T$ iff in the 
commutative limit they are invariant.
For the fermion kinetic term these statements are
a straighforward consequence of $\int {\widehat{\Psi_{\!L}}}^\dagger\*
{\fmslashmio\, \partial}{\widehat{\Psi_{\!L}}}
=\int {\widehat{\Psi_{\!L}}}^\dagger
{\fmslashmio \,\partial}{\widehat{\Psi_{\!L}}}$.
Since $\Fh$ transforms like $F$ under $C_{\!}P$ and $T$, and in the
$+\th$ case also under $C$ and $P$ separately,
the $C,P$,$T$ properties of the gauge kinetic term 
$\int T\!r(\Fh\*\Fh)=\int T\!r(\Fh\Fh)$ easily follow.
Inspection of the fermion gauge bosons interaction term
leads also to the same conclusion.

We have studied the $C,P$ and $T$ symmetry properties of NCYM 
actions where $\th$ transforms under $C,P$ and $T$ 
as a field strenght.
Viceversa, {\sl if  we keep $\th$ fixed} under $C,P$ and $T$ 
transformations, we in general have that NCYM theories 
{\sl break}  $C,P$ and $T$ symmetries. 

Finally a $U(1)$ 
field strenght is invariant under the combined $CPT$ transformation,
 and therefore $\th$ does not change. 
This implies that $CPT$ is always a  
symmetry of NCYM actions.

\sk
\noindent{\bf Acknowledgments$~$}
It is a pleasure to thank the organizers for the nice and 
stimulating
atmosphere at the conference and its efficient organization.


\end{document}